\documentstyle[11pt,newpasp,twoside]{article}
\def\edcomment#1{\iffalse\marginpar{\raggedright\sl#1\/}\else\relax\fi}
\reversemarginpar

\begin{document}
\title{Magnetic flaring in pre-main sequence stars}
\author{Eric D. Feigelson}
\affil{Department of Astronomy \& Astrophysics, Pennsylvania State
University, University Park PA 16802 USA}

\begin{abstract}
Observations of nearby star forming clouds with imaging X-ray
telescopes have revealed that X-ray emission is elevated $10^1-10^4$
above main sequence levels in low-mass pre-main sequence (PMS) stars.
The variability and spectral X-ray evidence, together with circularly
polarized radio continuum flares seen in a few cases, strongly argues
for an origin in magnetic reconnection flares.  These high levels of
magnetic activity are present from the earliest protostellar phase to
the main sequence.  After a brief review of past observations, three
astrophysical issues are raised: the location of the flaring magnetic
fields, the origin of these fields, and the effects of flare
high-energy photons and particles on the environs.  New results from
Chandra observations of a well-defined samples of PMS solar analogues
are presented, giving an improved view of magnetic flaring in the early
Sun.
\end{abstract}

\section{Introduction}

In his discovery paper of low-mass pre-main sequence (PMS) stars among
the Taurus-Auriga dark clouds, the T Tauri variables, Albert Joy (1945)
drew a connection between their unusual spectra and the solar
chromosphere.  In the following decades, hundreds of "flash variable"
low-mass stars were reported in Orion and other star-forming regions
(Haro \& Chavira 1966).  But study of magnetic activity played a
relatively minor role in the interpretation of young stellar objects
compared to the attention paid to emission lines associated with
outflows into and inflows from the circumstellar environment (e.g.
Herbig 1962; Hartmann 1998).

The presence of high levels of magnetic activity and flaring was
reasserted with the discovery of bright and highly variable X-ray
emission from PMS stars obtained with imaging X-ray telescopes.  These
X-ray luminous stars included both classical T Tauri stars and
similarly young stars without broad emission line spectra (Feigelson \&
DeCampli 1981; Feigelson \& Kriss 1981).  The latter "weak-lined" T
Tauri (WTT) stars are late-type, lithium-rich stars displaying a
variety of manifestations of enhanced magnetic activity: variable X-ray
emission $10^1-10^4$ times levels typically seen in older main sequence
stars; cool starspots covering large fractions of the stellar surface
evidenced by photometric rotational modulations or spectroscopic
Doppler imaging; variable H$\alpha$ and Mg II in emission from plage
and chromosphere; and, in some cases, U-band photometric flares,
variable circularly polarized radio continuum flares, and Zeeman
splitting of photospheric absorption lines.  WTT stars are sometimes
clustered around star forming molecular clouds along with CTT stars,
and sometimes dispersed tens of parsecs from older star formation
events such as the Sco-Cen OB Association and the Gould Belt.  These
findings are detailed in a review by Feigelson \& Montmerle (1999).

Here we examine in some detail one aspect of the magnetic activity of
young low-mass stars: the violent magnetic reconnection flares.  We
start with portraits of two flaring T Tauri and protostars from X-ray
and radio studies (\S 2).  Three astrophysical issues arising from such
observations are outlined (\S 3).  We then turn to an ongoing study of
a well-defined sample  of PMS solar analogues using the recently
launched Chandra X-ray Observatory (\S 4).

\section{Portraits  of two flaring PMS  stars: V773 Tau and YLW 15}

Perhaps the most magnetically active star in the nearby Taurus-Auriga
star forming cloud complex, V773 Tau = HD 283447 is a hierarchical
triple system of three roughly solar-mass stars with age $\simeq 1$
Myr. The system appears to be transitional between the CTT and WTT
phases. Its "quiescent" X-ray emission is very high $\simeq 1 \times
10^{31}$ erg/s in the soft X-ray band (0.5-2.4 keV), about $10^4$ times
the level of the contemporary Sun, with log$L_x/L_{bol} \simeq -3$ at
the "saturation  level" of late-type stars.

The ASCA satellite provided detailed views of two X-ray flares from
V773 Tau during 1995.  One event showed a modest elevation in flux in
the soft X-ray band ($0.5-2$ keV) but a factor $>50$ rise in the hard
band ($2- 8$ keV) followed by an exponential drop in flux with a decay
timescale of 2.3 hours (Tsuboi et al.\ 1998).  The inferred plasma
temperature at the flare peak was thus extremely high, $T(peak) \simeq
100$ MK with log$L_x(peak) \simeq 33.0$ erg/s and total energy release
log$E_x \simeq 37$ ergs in the X-ray band.  The decay phase is
reasonably well-modeled by simple quasi-static cooling of a uniform
plasma where the volume emission measure $EM \propto T^3$.  For this
model, the electron density is inferred to be $n_e \simeq 3 \times
10^{11}$ cm$^{-3}$ with a loop length $\simeq  4 \times 10^{11}$ cm
$\simeq 2$ R$_*$.  The second ASCA flare had  somewhat lower peak
emission with $T(peak) \simeq 40$ MK and log$L_x(peak) \simeq 32.3$
erg/s, but a longer flux decay over $>20$ hours (Skinner et
al.\ 1997).  A short-lived reheating (or separate flare) event was seen
during the decay.  The temperature and flux decay phase can be roughly
modeled by either a quasi-static cooling loop or a two-ribbon
long-duration solar flare model.  Another possibility is rotational
eclipsing of a continuously emitting loop near the rotational pole with
height around 0.6 R$_*$.

Both members of the V773 Tau close binary are strong nonthermal radio
emitters, as directly imaged with VLBI techniques.  On one occasion, a
radio flare was seen with the Very Large Array showing both linear and
circular polarization, indicating acceleration of  electrons to
supra-relativistic energies during the flare (Phillips et al.\ 1996).
Radio and X-ray flaring are decoupled from each other.  While V773 Tau
is particularly well-studied,  WTT stars with similar extremely high
magnetic activity include V410 Tau and HDE 283572 in Taurus-Auriga,
DoAr 21 in Ophiuchus, and Parenago 1724 in Orion.  Simple models of
single magnetic loops consistently suggest enormous loop size of
several stellar radii.

The molecular cores of the nearby $\rho$ Ophiuchi cloud is the richest
cluster of young stellar objects within $\simeq 200$ pc of the Sun.  In
addition to dozens of T Tauri stars are several embedded Class I
protostars. YLW 15 = IRS 43 is one of the youngest and most luminous of
these protostars with $L_{bol} \simeq 10$ L$_\odot$ and estimate age of
$\simeq 0.1$ Myr.  It is surrounded by a dense dusty envelope $\simeq
3000$ AU in radius and produces a small bipolar CO outflow with an
ionized region around the protostar seen in radio continuum.

Its magnetic activity was first detected in soft X-rays during an
intense flare with the ROSAT satellite, rising by a factor $>20$ and
decaying with an $e$-folding time of 5 hours  (Grosso et al.\ 1997).  A
long ASCA exposure revealed a  sequence of three flares separated by
$\simeq 20$ hours (Tsuboi et al.\  2000).  Each showed a fast rise to a
peak log$L_x \sim 32$ erg/s, followed by an exponential decay with
timescales of $8-18$ hours. The plasma cooled during the decays from a
peak $T \simeq 60$ MK.  The total energy of the flare sequence in the
X-ray band was log$E_x \simeq 37.0$ ergs.  As with the WTT stars,
interpretation  based on radiatively cooling loops imply loop lengths
$l \simeq 1 \times 10^{12}$ cm and plasma densities comparable to those
in solar flare ($n_e \sim 10^{10}$ cm$^{-3}$).  Montmerle et
al.\ (2000) develop the idea that the reconnection field lines connect
the protostar to the circumstellar disk near the star-disk corotation
radius;  such field lines will be continuously twisted by the Keplerian
velocity shear in the disk.  The model parameters indicate a  protostar
with mass $M_* \simeq 2$ $M_\odot$ and radius $R_* \simeq 5$ R$_\odot$
rotating rapidly with period $P_* \simeq 10$ hours linked to the inner
disk at $R_{disk} \simeq 2$ R$_*$ with a much slower orbital period
$P_{disk} \simeq 4$ days.  They suggest  that YLW 15 is in the early
stages of star-disk magnetic braking.  Radio continuum emission seen
from the system is interpreted as thermal emission from
coronal-temperature plasma ejected by the reconnection events, not
nonthermal emission as in V773 Tau.

\section{Three astrophysical issues}

\subsection{What is the magnetic field geometry producing young stellar
flares?}

The basic concept of angular momentum transfer from the protostar to
the disk via star-disk magnetic fields (K\"onigl 1991), adapted from
interactions between neutron stars and their accretion disks, is widely
accepted, and the theory for accretion and generation of collimated
outflows from the corotation radius is well-developed (e.g.\ Shu et
al.\ 1994).  Considerable empirical evidence supports this star-disk
magnetic interaction model including interpretation of optical emission
lines and outflow properties for T Tauri stars (e.g.  Hartmann 1998;
Feigelson \& Montmerle 1999; Guenther, this volume).  

The question is whether these star-disk fields are responsible for the
magnetic flaring seen with X-ray and radio telescopes, as in the model
of YLW 15 outlined above.  Even after magnetic braking is largely
complete, fluctuations in accretion plausibly can lead to field
reconnection, as indicated in magnetohydrodynamical simulations.  The
high $L_x(peak)$ and $L_r(peak)$ luminosities together with the long
durations of PMS flares compel extremely large magnetic structures that
extend far beyond the stellar surface if one assumes a simple model of
an isolated magnetic loop with a single injection of magnetic energy
undergoing radiative cooling.

The uncertainty in this model arises because long-duration flares are
seen in the Sun and, with comparable luminosities, in some RS CVn-type
magnetically active stars which do not involve links to a circumstellar
disk.  In long-duration solar flares which last many hours, energy is
stored in a complex of magnetic structures which are released
sequentially.  In some cases a single loop is repeatedly reheated,
while in other cases motion or eruption of one structure causes a flare
in a nearby structure.  The flares of WTT stars, which presumably are
no longer interacting with a disk,  are similar in duration and power
to protostellar flares.  As there is a broad consensus that WTT
activity is due to solar-type flaring in multipolar structures attached
to the stellar surface, it is plausible that this is the only geometry
producing flares in even the youngest PMS stars.  (See the Discussion
below for additional comments on this issue.)

\subsection{Does PMS magnetic activity arise from a solar-type dynamo?}

There are profound reasons, both astrophysical and observational, to
believe that a magnetic dynamo is the source of surface activity on the
Sun and in main sequence late-type stars (Cattaneo, Dikpati, Knoelker,
Charbonneau, this volume).  The same processes should be present in PMS
stars which have mostly or fully convective interiors and are rotating
more rapidly than most main sequence stars.  It seems implausible that
PMS stellar interiors rotate as perfect solid bodies to avoid the
twisting and amplification of fields which generate stellar dynamos.

Yet, the few available observational diagnostics of the origin of PMS
activity provide a muddled picture and do not clearly support the
dynamo model.  The relationship between X-ray emission (measured either
by X-ray luminosity or X-ray surface flux) and stellar rotation
(measured either by spectroscopic surface velocities or photometric
rotational periods $P_*$, or Rossby number) is either weak or absent in
available samples.  In particular, slowly rotating PMS stars are often
seen at very high X-ray levels (Figure 1).  This dramatically contrasts
with main sequence stars where, for solar-mass stars in the range $3 <
P_* < 30$ days, X-ray luminosity is anti-correlated with rotational
period according to log$L_x \simeq 31.0 - 2.6 \log P_*$ (G\"udel et
al.\ 1997).  Instead of a $L_x - P_*$ association, the principal
correlations seen in PMS samples are $L_x - L_{bol}$ (characteristic of
OB stars, not late-type, main sequence stars) and $L_x - M_*$
(Feigelson et al.\ 1993).  The situation is poorly understand, as it is
difficult to untangle interwovern effects of luminosity, age, mass and
rotation in PMS stars.

\begin{figure}[!ht]
\plotfiddle{"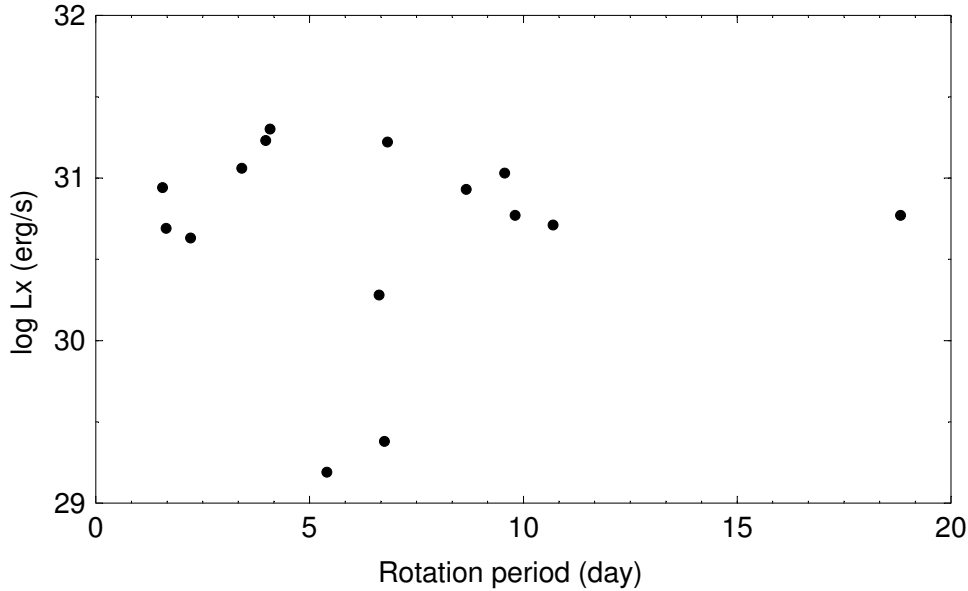"}{3.0in}{0.0}{80.}{80.}{-250.0}{-330.0}
\caption{Relationship between X-ray luminosity $L_x$ ($0.5-8$ keV band)
and rotational period $P_*$ measured from photometric modulation of
starspots for PMS analogues of the Sun.   It does not show the
anti-correlation seen in main sequence stars and expected from the
magnetic dynamo model.  The sample shown here is discussed in \S 4.
{\it (Feigelson et al., in preparation)}}
\end{figure}

\subsection{What are the effects of flares on the circumstellar
environment?}

Some of the most important astrophysical implications of PMS flaring
may involve the impact of flare products on ambient material.  These
include keV-energy photons as seen with X-ray telescopes, MeV-energy
particles as evidenced by gyrosynchrotron emission in PMS stars, and
shocks that are likely to accompany the violent reconnection events.
These effects are complex and, in most cases, their role is quite
uncertain.  These issues are discussed in some detail by Glassgold et
al.\ (2000).

PMS X-rays, both during flares and quiescence, must impact some part of
the circumstellar disks and partially ionize the largely neutral
material.  As the geometry of the X-ray emitting magnetic structures is
uncertain (\S 3.1), it is unclear whether large portions of the disk
will in fact be irradiated.  There is some empirical evidence for X-ray
irradiation of the disk:  a 6.4 keV fluorescent emission line from
neutral iron seen in flare of protostar R1 Cr A (Koyama et al.\ 1996),
and excited molecular lines in the disk of TW Hya (Weintraub et
al.\ 2000).  It is possible that this ionization will induce
magnetohydrodynamical behavior, such as the magneto-rotational (Balbus-
Hawley) instability, which in turn would stimulate turbulent viscosity,
accelerate angular momentum transport within the disk, and promote
accretion onto the young star.  Ionization of the outer layers of the
disk may also be critical for coupling the Keplerian rotation of disk
material to the collimated outflows characteristic of protostars and
CTT stars.

For deeply embedded PMS stars, X-rays must ionize some molecules in the
surrounding cloud.  If these X-ray Dissociation Regions (Hollenbach \&
Tielens 1997) extend over sufficiently high over sufficiently large
volumes of the cloud, then ambipolar diffusion of the neutral gas
thrugh cloud magnetic fields will be impeded and subsequent star
formation in the cloud may be delayed.  There is as yet little
astronomical evidence yet for XDRs around flaring PMS stars.

Gyrosynchrotron emission from MeV electrons have been seen in dozens of
flaring PMS stars, mostly WTT stars but in one case a Class I protostar
(Feigelson \& Montmerle 1999).  If the acceleration processes are
similar to those seen in solar flares, relativistic protons and other
nuclei will also be produced which will bombard some portion of the
disk.  There is good evidence that this process occurred in the early
solar nebula: certain grains in carbonaceous chondrites
exhibit both unusually high levels of particle tracks and spallogenic
$^{21}$Ne which can only be plausibly explained by exposure to very
high MeV particle fluences prior to compaction into the host meteorite
(Woolum \& Hohenberg 1993).  With less confidence and more controversy,
PMS flare protons may also account for some or all of the remarkable
high abundances of short-lived nuclides (e.g.  $^{26}$Al, $^{41}$Ca)
found in carbonaceous chondrites (Feigelson 1982; Lee et al.\ 1998).

Finally, the shocks and radiation produced by PMS flares may provide
the sudden heating that melted dustballs in the solar nebula, producing
the chondrules found in great abundance in stony meteorites.  Shu et
al.\ (1997) develop such a scenario that accounts for a variety of
chondrules properties, including mass sorting in different meteorites.
Many  other models for chondrules melting have been suggested, and as
yet none has gained widespread acceptance.

\section{Chandra study of solar analogues in the Orion Nebula Cluster}

The Chandra X-ray Observatory, launched in July 1999, has a unique
combination of capabilities for the study of X-ray emission from young
stars, particularly those that are in young clusters still embedded in
their molecular cloud.  Its superb mirrors provide arcsecond imaging
and reflectivity up to $\simeq 8$ keV; the CCD detectors of the ACIS
detector provide extremely low noise, high quantum efficiency and
moderate spectral resolution; and the high-Earth orbit permits long
uninterrupted observations of variable phenomena.  Young stars can be
detected with as few as 7 counts during day-long exposures, and
embedded objects are seen with absorptions up to $A_V \simeq 50-100$
magnitudes.

A prime Chandra target is the Orion Nebula Cluster (ONC), a dense
cluster of $\simeq 2000$ stars with masses ranging from 0.02 to 50
M$_\odot$ and ages from $\simeq 0.1$ to $\simeq 10$ Myr which
illuminates the famous Orion Nebula.  Early findings are reported by
Garmire et al.\ (2000) and Schulz et al.\ (2001).  Nearly all of the $M
\geq 1$ M$_\odot$, and a large fraction of the lower-mass stars, are
detected with ACIS -- this image with $\simeq 1000$ X-ray stars is the
richest field of sources ever obtained in X-ray astronomy.

Here I present preliminary results concerning a well-defined subsample
of ONC stars that can be considered analogues of the early Sun
(Feigelson et al.,  in preparation).  It consists of all ONC stars with
$V<20$ and $0.7 < M < 1.4$ M$_\odot$.  Thirty-nine of these 41 stars
are detected with ACIS; most are extremely strong with $>1000$ photons
detected in two $\simeq 12$ hr exposures separated by several months.
Luminosities are measured from plasma fits to the ACIS spectra in the
$0.5-8$ keV band assuming a distance of 450 pc.

A wide range of variability characteristics are seen among these
solar-mass PMS stars:  some show constant emission while about 1/3
exhibit powerful short-timescale flares during the two 12-hour
observations.  These flares have a wide range of properties:  $30 <
\log L_x(peak) < 32$ erg/s; $34 < \log E_x < 36$ erg, and durations
from 40 minutes to $>12$ hours. Most, but not all, show spectral
hardening around peak flux and softening during the decay phase.
Prominent flares are seen in both CTT and WTT solar analogues, and in
stars of all ages from $\sim 0.1$ to $\sim 20$ Myr.  Figure 2 shows
three of these flares.  The top panel is an extremely young WTT star
with flare showing log$L_x(peak) \simeq 32.0$ erg/s and $E_x > 36.2$
erg.  The middle panel is an intermediate-age CTT star with flare
showing log$L_x(peak) \simeq 31.4$ erg/s and $E_x \simeq 34.7$ erg.
The bottom panel is an older CTT star with flare showing
log$L_x(peak)>31.5$ erg/s and $E_x > 35.7$ erg.  CTT (classical T
Tauri) and WTT (weak-lined T Tauri) classification is based on K-band
infrared excess.  No obvious relationship between flare characteristics
and stellar age, rotation or disk emission is apparent.

\begin{figure}[!ht]
\plotfiddle{"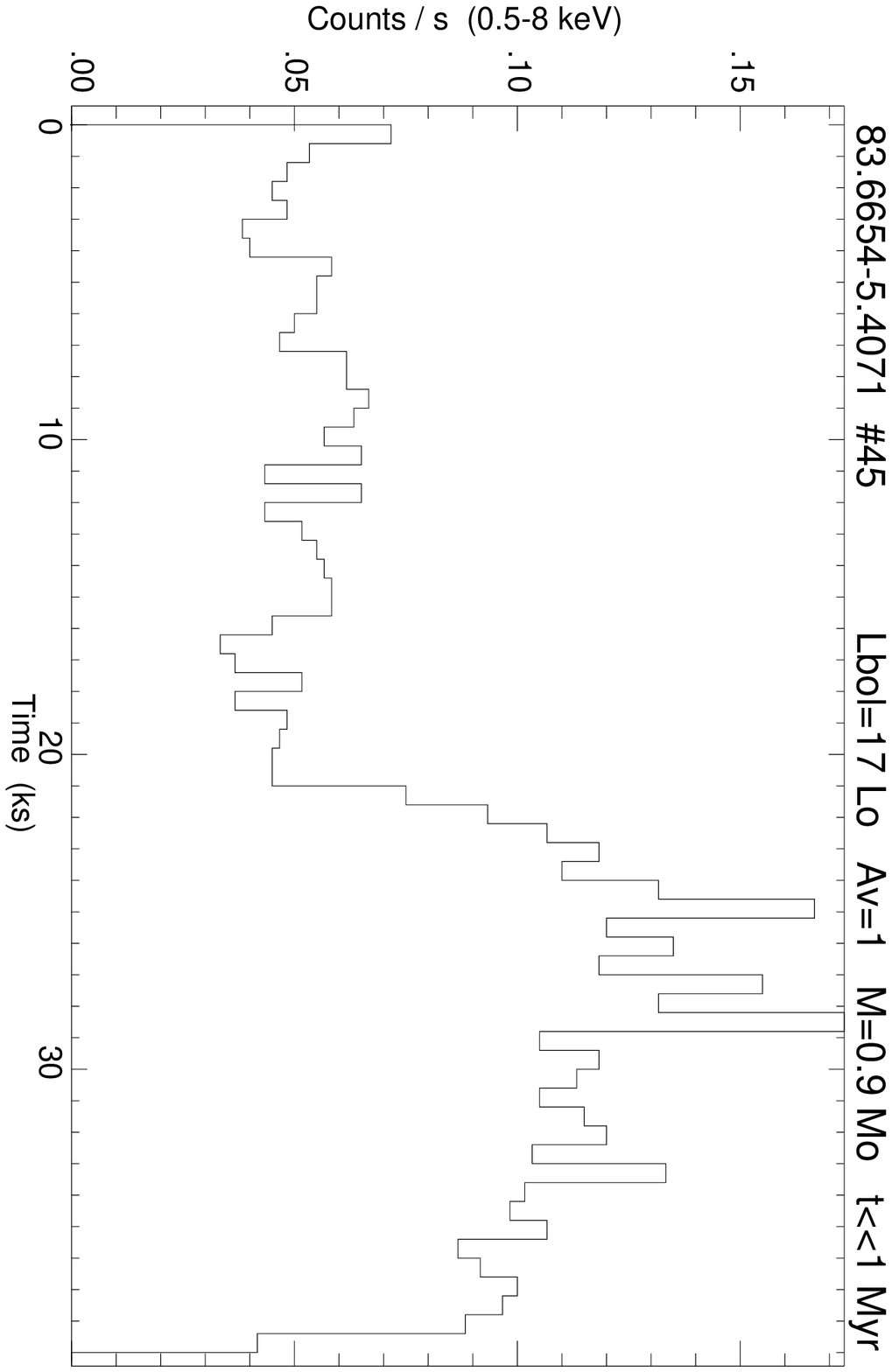"}{2.3in}{90.0}{45.}{45.}{200.0}{-30.0}
\plotfiddle{"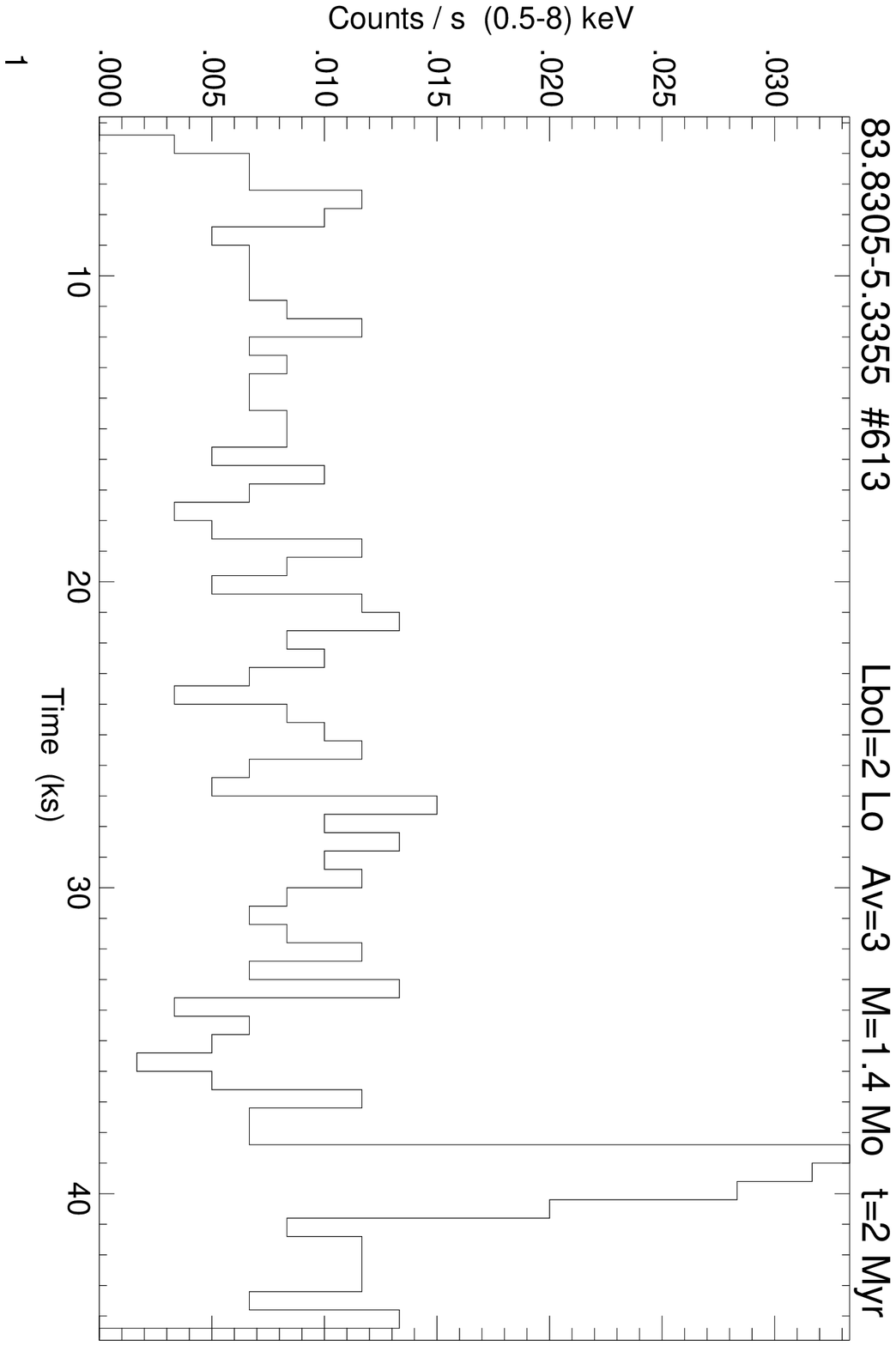"}{2.3in}{90.0}{45.}{45.}{200.0}{-30.0}
\plotfiddle{"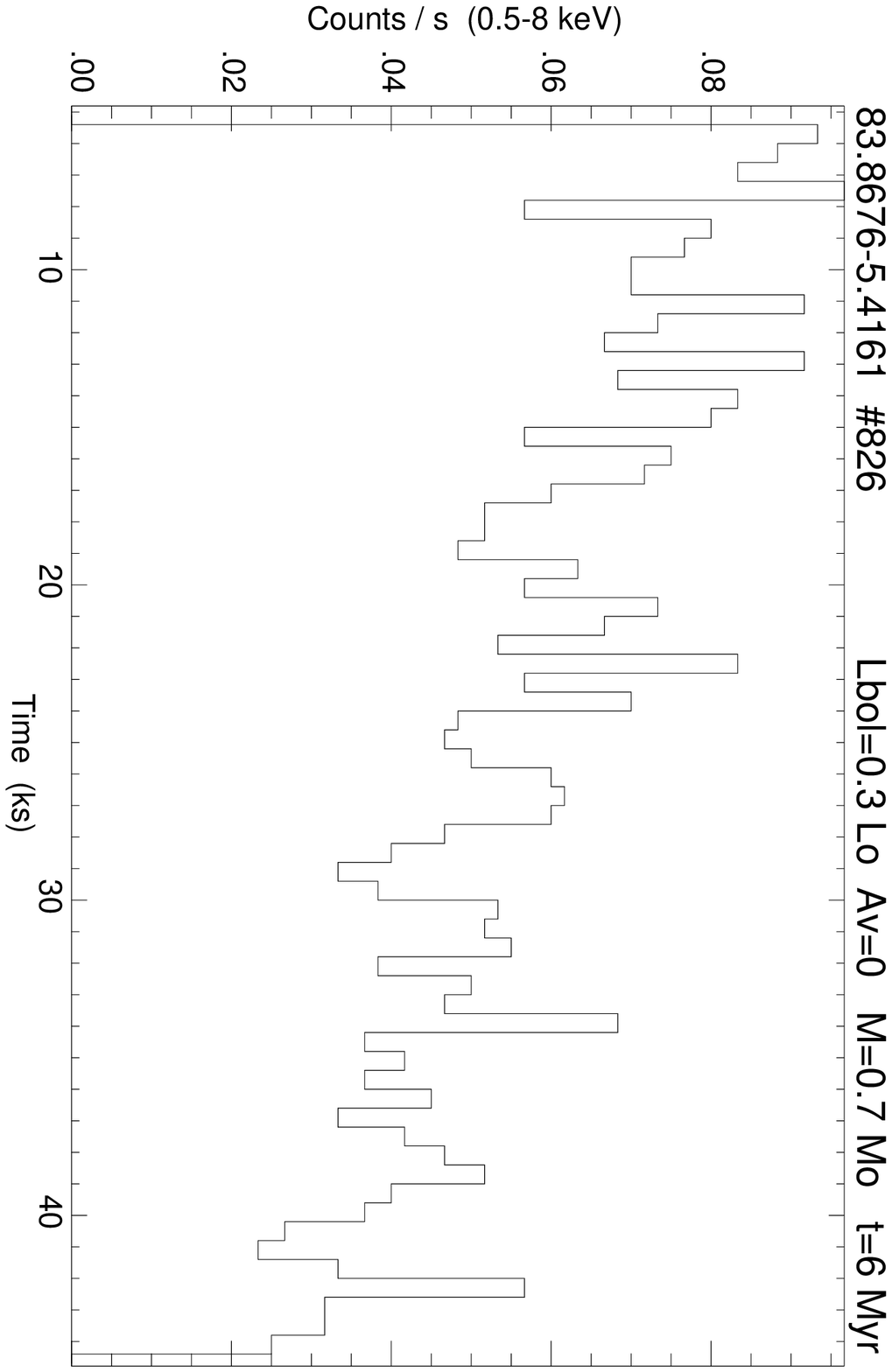"}{2.3in}{90.0}{45.}{45.}{200.0}{-30.0}
\vspace{-0.1in}
\caption{X-ray lightcurves from the Chandra X-ray Observatory
observations Orion Nebula Cluster, illustrating the range of flares
present in PMS solar analogues.  Each panel is labeled by the stellar
counterpart, bolometric luminosity, visual absorption, mass and age.
The duration of each light curve is about 12 hours. 
{\it (Feigelson et al., in preparation)}}
\end{figure}

The time-averaged X-ray emission of the sample declines with stellar
age: log$L_x$ falls from about 31 to $29-30$ erg/s over a few Myr.
This can be readily explained as a combination of magnetic saturation,
where virtually the entire surface of the star is covered with magnetic
structures, and the declining surface area as the star evolves down the
convective Hayashi track. These very young stars all have
log$L_x/L_{bol} = -3.5 \pm 0.5$, near the upper envelope of
$L_x/L_{bol}$ seen in magnetically active main sequence stars.  After
$1-2$ Myr, the dispersion in X-ray emission appears to increase: some
stars remain with log$L_x/L_{bol} \simeq -3.5$ (indeed, a few become
even more active with log$L_x/L_{bol} \simeq -2$), while in others the
X-ray emission declines to log$L_x \leq 29$ erg/s and log$L_x/L_{bol}
\leq -4.5$.  This phenomenon might be understood in the context of
models of the rotational evolution of PMS stars (Bouvier et al.\ 1997)
combined with simple dynamo concepts.  Stars which remain rotationally
coupled during the descent of the Hayashi track will be slow rotators
and, would display low levels of magnetic activity like X-ray
luminosity, while stars that decouple early from their disks will spin
up as they contract and display high X-ray levels. However, we do not
see the anti-correlations between circumstellar disks or rotational
periods and X-ray emission in these solar-mass PMS stars (Figure 1).
The situation is thus confused, and the data conceivably may falsify
the magnetic dynamo model along the PMS convective tracks.

Importantly, since our sample is complete, we can be fairly confident
that the behavior seen in ONC solar analogues represents behavior
exhibited by the early Sun.  Several consequences of this solar-stellar
link emerge.  First, during the first $1-2$ Myr, the surface of the
young Sun was  "saturated" with magnetic fields which heated plasma to
temperatures $\geq 10$ MK emitting $0.5-8$ keV X-rays at levels of
log$L_x \simeq 31$ erg/s and log$L_x/L_{bol} \simeq -3.5$.  Afterwards,
during the $2-20$ Myr period, we cannot determine if the Sun's activity
remained at high levels or decreased by $1-2$ orders of magnitude.
During the entire PMS era, the early Sun produced a flare at levels
log$L_x \geq 30$ erg/s and log$E_x \geq 34$ erg every few days. For
comparison, the contemporary Sun produces a long-duration flare $10^2$
weaker roughly every $10^3$ days (viz. the 1992 Nov 2 flare).

These results provide the first {\it quantitative} determination of the
activity and flaring levels in the early Sun based on astronomical
measurements of a well-defined sample of young solar analogues.  They
can be used to evaluate the importance of local high energy photons and
particles on the solar nebula: ionization of gas, charging of dust
grains, spallation of nuclei in meteoritic solids, and flash melting of
chondrules.

{\it Acknowledgements:}  This work was supported by NASA contract 
NAS8-38252 and grant NAG5-8422.

\newpage
\section*{Discussion}

\noindent PRIEST: It is worth examining whether solar flares can be
scaled to give the observed T Tauri flare properties (see E.\ Priest
and T.\ Forbes, {\it Magnetic Reconnection}, Cambridge Univ Press,
2000).  For example, $10^{37}$ ergs seem enormous to a solar physicist,
but in the Sun the energy scales as $B^2 l^3$ so that field strengths
($B$) and sizes ($l$) that are both a factor of 10 higher
would produce the PMS flare.  Also, on the Sun, the `decay'
time is certainly not the radiative cooling time.  It has proved very
difficult to build a realistic solar flare model: having X-points and
reconnection is not sufficient since reconnection can occur
continuously, and the presence of an instability is not sufficient
since it may saturate.  One needs to demonstrate how energy can slowly
build up and then suddenly be released, as in the magnetic catastrophe
model.  So perhaps some of the solar flare experience can help with
understanding your PMS flares.

\noindent SCHMITT:  On the issue of the location of CTT flares, in a
flare on Algol with $E_{bol} > 10^{37}$ ergs similar to CTT stellar
flares, the X-ray emission was eclipsed (Schmitt \& Favata, Nature
401:44, 1999).  This shows the flaring plasma to be near the pole and
with limited height.  The flare energetics can be explained by
reconnection of magnetic fields of approximately 1 kG in the corona.
Thus, in my opinion, there is no need to invoke star-disk interactions
in CTT stars.

\noindent FEIGELSON:  Both of these points are cogent criticisms of the
frequent use of the over-simplified quasi-static cooling loop model for
PMS stellar flares.  Dr. Schmitt's argument is discussed in more detail
by F. Favata (in {\it X-ray Astronomy 2000}, R. Giacconi et al.\ eds,
in press).  It is difficult to discriminate between large star-disk and
compact star-star loop flares from the existing data.  There is hope
that more detailed spectral studies (e.g.\ of the fluorescent
Fe-K$\alpha$ 6.4 keV line from disk reflection) may provide critical
clues.

\noindent LINSKY:  You asked whether PMS flares occur close to or far
from the star.  One test is to study the change in metal abundance: for
RS CVn systems, metal abundances increase during flares presumably due
to evaporation of material from the photosphere or chromosphere.  Can
this be done for PMS flares?

\noindent FEIGELSON:  A valuable thought, particularly in light of the
dramatic abundance variations during stellar flares found with XMM
(Brinkman et al., As\&Ap 365:L324, 2001).  With ACIS alone, it is
difficult to make detailed statements about individual elemental
abundances, but abundance study of a bright PMS stars with the Chandra
or XMM gratings is definitely warranted.

\noindent VAN  BALLEGOOIJEN:  Are there simultaneous observations of
X-ray and H$\alpha$ flares in pre-main sequence stars?

\noindent FEIGELSON:  Simultaneous X-ray, optical/UV and radio
measurements have been made for a few stars (e.g., Feigelson et
al.\ ApJ 432:373, 1994;  Guenther et al.\ As\&Ap 357:206 2000).
Generally there is no relation between X-ray and radio flares, but a
weak correlation between X-rays and H$\alpha$ may be present.

\noindent PISKUNOV:  In the case of long duration soft X-ray events in
CTT stars, how can you tell a flare from variability in the mass
accretion rate?

\noindent FEIGELSON: It is doubtful that a significant fraction of
observed X-rays comes from accretion, as the observed X-ray
temperatures are typically $10-30$ MK (and higher during powerful
flares) compared to $\sim 1$ MK expected from the accretion shock
(e.g.\ Lamzin et al., As\&Ap 306, 877, 1996).

\end{document}